\documentclass{optica-article}

\journal{opticajournal} 

\articletype{Research Article}
\usepackage{amsmath}
\usepackage{amsfonts}
\usepackage{lineno}
\usepackage{xcolor}

\begin{document}

\title{Polarization Vortex for Enhanced 
Refractive Index Sensing}
\author{Ravshanjon Nazarov,\authormark{1,2} Mingzhao Song,\authormark{1} Andrey Bogdanov,\authormark{1,2*}  and Zarina Kondratenko\authormark{2*}}

\address{\authormark{1}Harbin Engineering University, Qingdao Innovation and Development Center, Sansha road 1777,
266000, Qingdao, Shandong, China\\
\authormark{2}School of Physics and Engineering, ITMO University, 197101, St. Petersburg, Russia}

\email{\authormark{*}a.bogdanov@metalab.ifmo.ru}
\email{\authormark{*}z.sadrieva@metalab.ifmo.ru} 
\

\begin{abstract*}Although all-dielectric sensors exhibit minimal absorption and a high figure of merit (FOM), their sensitivity is significantly lower compared to plasmonic sensors. One approach to enhancing the sensitivity of dielectric sensors is utilizing bound states in the continuum (BICs), which are resonant states with an infinite radiative lifetime.~These states are characterized by polarization vortices in the far field, whose winding number
determines the topological charge. Here, we demonstrate that the position of a BIC polarization vortex in the $k$-space has a square-root dependence on changes in the refractive index of the medium similar to an exceptional point. We compute the angular and spectral sensitivities of our structure and demonstrate that the angular sensitivity reaches values comparable to those of surface plasmon polariton (SPP)-based sensors.~We observe a blue spectral shift of BICs as the refractive index of the surrounding medium increases, a behavior that differs from the conventional spectral response typically expected under such perturbations.~Additionally, we found a distinct BIC regime exhibiting a pronounced angular sensitivity, surpassing its spectral one. Our findings pave the way for the development of dielectric sensors with high angular sensitivity and facilitate the practical observation of the optical vortex dynamics.

\end{abstract*}
\section{Introduction}

Refractive index sensing allows detecting small changes in the optical properties of materials. This capability is essential for a wide range of applications, including biochemical sensing, environmental monitoring, and medical diagnostics~\cite{pitruzzello2018photonic}. One of the traditional approaches is based on surface plasmon polariton resonances (SPP-sensors), that provide sensing by detecting shifts in the angle of incidence at a fixed excitation frequency~\cite{kabashin2009phase}. Although they are relatively affordable, these methods suffer from a limited figure of merit (FOM) due to substantial material losses~\cite{kravets2018plasmonic}. Similarly, due to technological constraints and sensitivity to external factors, fiber-optic sensors often lack the ultra-high sensitivity required to detect slight changes in the refractive index~\cite{zhang2024distributed}.~To overcome these drawbacks, recent efforts have focused on exploring resonant states with infinitely long lifetimes, robust against structural perturbations~\cite{romano2018label,schiattarella2022high}. These unique states, known as bound states in the continuum (BICs), have attracted significant attention in recent years due to their remarkable properties, including highly sensitive and efficient sensing abilities.

Bound states in the continuum (BICs) are non-radiating resonant states whose energy spectrum is embedded in the continuum spectrum of propagating waves.~Being high-$Q$ states, BICs were shown to enhance light-matter interaction~\cite{mocella2015giant,yoon2015critical}, sensors~\cite{pitruzzello2018photonic}, lasing~\cite{kodigala2017lasing,huang2020ultrafast,liu2023nonlinear}, second harmonic generation~\cite{zograf2022high,bernhardt2020quasi}, filtering of light~\cite{cui2016normal,foley2014symmetry} and enable manipulation of phase singularities~\cite{liu2021evolution,valero2023exceptional}. 
In experimental spectra, BICs are manifested 
as a collapse of the Fano resonance when the width of the resonance vanishes~\cite{bogdanov2019bound}. This allows the creation of sensors sensitive to small perturbations in the optical parameters (Ref.~\cite{maksimov2020refractive,liu2017optical,wang2017optofluidic}). Despite these advantages, controlling BICs in practice is challenging: it requires precise adjustment 
of structural parameters---a task that is hard to achieve with 
nano-fabrication techniques~\cite{xiao2022manipulation}. 
To overcome these fabrication drawbacks, a possible solution is to create BICs with high topological charges, which can increase their robustness and stability~\cite{kang2022merging}.

In the far field, BICs can be observed as polarization vortices, with their winding numbers determining the topological charge.~The presence of a topological charge characterizes the robustness of BICs against perturbations, as long as the system retains its in-plane symmetry~\cite{zhen2014topological}. The topological nature of BICs was explored in one-dimensional lattices~\cite{doeleman2018experimental}, arrays of spheres and discs~\cite{bernhardt2020quasi,sidorenko2021observation,ochiai2024generation}, metasurfaces~\cite{jin2019topologically,bernhardt2020quasi}, and other structures~\cite{arjas2024high}. Namely, recent studies have shown that under a symmetric perturbation, BICs do not disappear; moreover, they can migrate in parametric space~\cite{zhen2014topological,jiang2023general,bulgakov2017bound,song2024evolution}. During this migration, BICs can be transformed or annihilated, resulting in intriguing phenomena.~For instance, merging of BICs can create 
 ultra-high $Q$ resonances~\cite{kang2022merging}, while a higher-order topological charge enhances robustness against structural defects, which is important for applications~\cite{zhen2014optical}. As shown in Refs.~\cite{qi2023steerable,song2024evolution,bulgakov2017bound}, BIC positions in the $k$-space can be controlled by 
 varying geometrical parameters of the structure. However, achieving these results in an experiment is challenging: a large number of samples should be fabricated to detect a vortex shift or transformation of BICs. Another method for manipulating the topological charge position in the parametric space is varying the structure's refractive index without modifying its symmetry. For instance, this can be realized through optical Kerr-type nonlinearity or by varying the refractive index of the medium~\cite{lv2020high}.

Here, we investigate the migration of
BICs in the $k$-space induced by variations in the optical contrast between the surrounding medium and the photonic structure.~We observe distinct blue and red shifts in the resonance frequency for specific BIC regimes under this perturbation. Moreover, we demonstrate that tracing the angular shifts caused by changes in the refractive index offers more sensitive detection compared to 
spectral shifts. We calculate the sensitivity of BIC positions to variations in the refractive index of the surrounding medium and show that a square-root dependence can be achieved—matching the sensitivity levels of sensors based on exceptional points (EPs)~\cite{wiersig2020review}. Using perturbation theory, we provide a theoretical explanation for these results. Additionally, we reveal BICs with remarkably high angular and spectral sensitivities, surpassing those of the photonic bands. 
This regime enables novel sensor designs, in which shifts
of vortices in the $k$-space due to refractive index changes become more informative than conventional resonance shifts. This
behavior paves the way for ultrasensitive optical sensing platforms leveraging the unique properties of topological photonic structures.

\section{BIC migration} 
 \begin{figure*}[htbp]
 \centering
\includegraphics[width=0.8\linewidth]{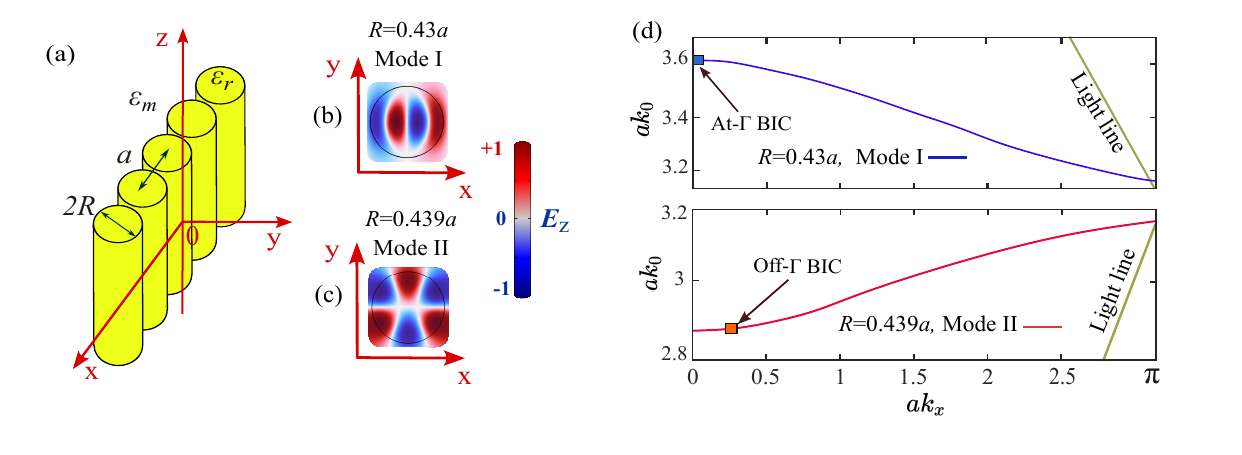}
\caption{(a) Geometry of the rod array with indicated parameters. (b-c) Electric field distribution at the resonance wavelengths for $\varepsilon_{m}=1$, $\varepsilon_{r}=15$, and two different values of $R$. (d) Band diagrams with indicated positions of BICs corresponding to the modes from panels (b,c), respectively.}
\label{fig:structure}
\end{figure*}
We consider two arrays of dielectric rods, with different radii $R$ and with a fixed permittivity $\varepsilon_r$. The rods are infinitely long in the $z$ direction and placed periodically in the $x$ direction with period $a$ in a medium with $\varepsilon_m$, see Fig.~\ref{fig:structure}(a). The radii of the rods are equal to $R=0.43a$ or $R=0.439a$, while the initial permittivities of the medium and the rods are $\varepsilon_m=1$ and $\varepsilon_r=15$, respectively. The array is illuminated by a TE-polarized plane wave at a frequency within the subdiffractive band, where only one diffraction channel is open. Figure~\ref{fig:structure}(d) shows the band diagram of the modes with indicated BIC positions, while the mode profiles $E_z$ at the resonance frequency are shown in Figs.~\ref{fig:structure}(b) and 1(c).  

\begin{figure} [b!]
  \centering
  \includegraphics[width=0.55\linewidth]{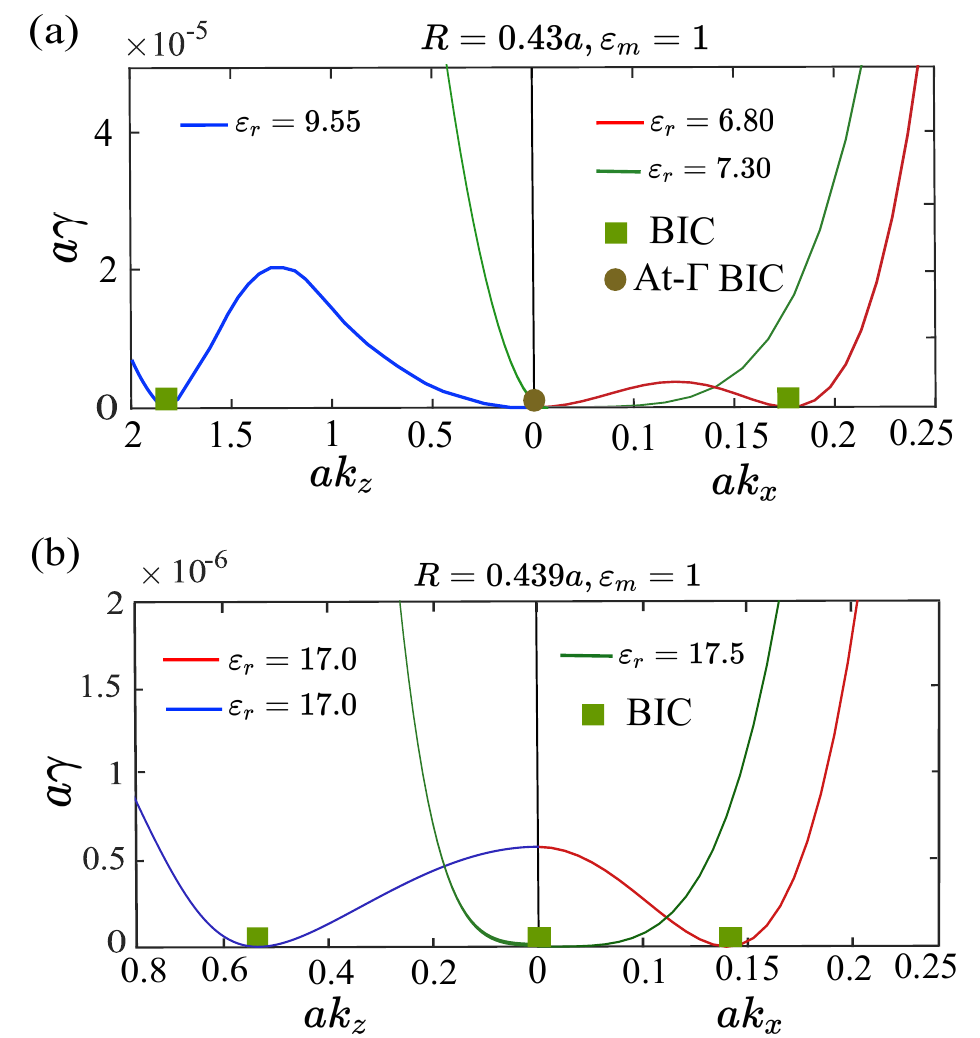}
  \caption{Dependence of the normalized radiative losses $a\gamma$ on the wavevector components for the rods with (a) $R=0.43a$ and (b) $R=0.439a$. BICs are indicated by green squares. With changes in the rod permittivity $\varepsilon_r$, BIC changes its position, i.e., migrates in parametric space. A circle denotes at-$\Gamma$ BIC that remains at the center of the zone when $\varepsilon_r$ changes.}
  \label{fig:eigen}
\end{figure}

The band diagram of the structure and BIC positions with mode profiles are obtained by means of COMSOL Multiphysics software. Since BICs have no radiative losses, i.e.~$\gamma=0$, one should select only real-valued solutions to find BICs. The results of the eigenvalue problem for several values of optical contrast (difference between the rod and medium permittivity) $\Delta\varepsilon=\varepsilon_r-\varepsilon_m$ are shown in Fig.~\ref{fig:eigen}, where the minima of the radiative losses correspond to BICs. Figure~\ref{fig:eigen} shows that as the optical contrast changes, these minima take on different values in the parametric space. In particular, for the array with $R=0.43a$, as
 $\varepsilon_r$ decreases, BICs migrate from the $k_z$ to $k_x$ direction. Meanwhile, the at-$\Gamma$ BIC remains in the same position~[see Fig.~\ref{fig:eigen}(a)]. The second array with $R=0.439a$ supports the migration of BICs to the $\Gamma$ point as $\varepsilon_r$ grows [Fig.~\ref{fig:eigen}(b)].

Since BICs are topologically protected states, we define topological charge through the far-field polarization as~\cite{zhen2014topological}:

\begin{equation}
\label{Tcharge}
    q=\frac{1}{2\pi}\oint_C \text{d} \textbf{k}\nabla \arg(\text{{E}}({\textbf{k}})),
\end{equation}
where the integral is taken along a closed path $C$ in the $k$-space that encircles the BIC in the counter-clockwise direction; $\text{E}({\textbf{k}})=c_{x}(\textbf{k}) + {i}c_{z}(\textbf{k})$ with $c_{x}(\textbf{k})$ and $c_{z}(\textbf{k})$ being
 the Cartesian components of the polarization vector. Therefore, to distinguish the topological charges, we need to calculate polarization maps in the far field and find the rotation of polarization ellipses. 
\begin{figure*}[t!]
             \centering
    \includegraphics[width=0.9\linewidth]{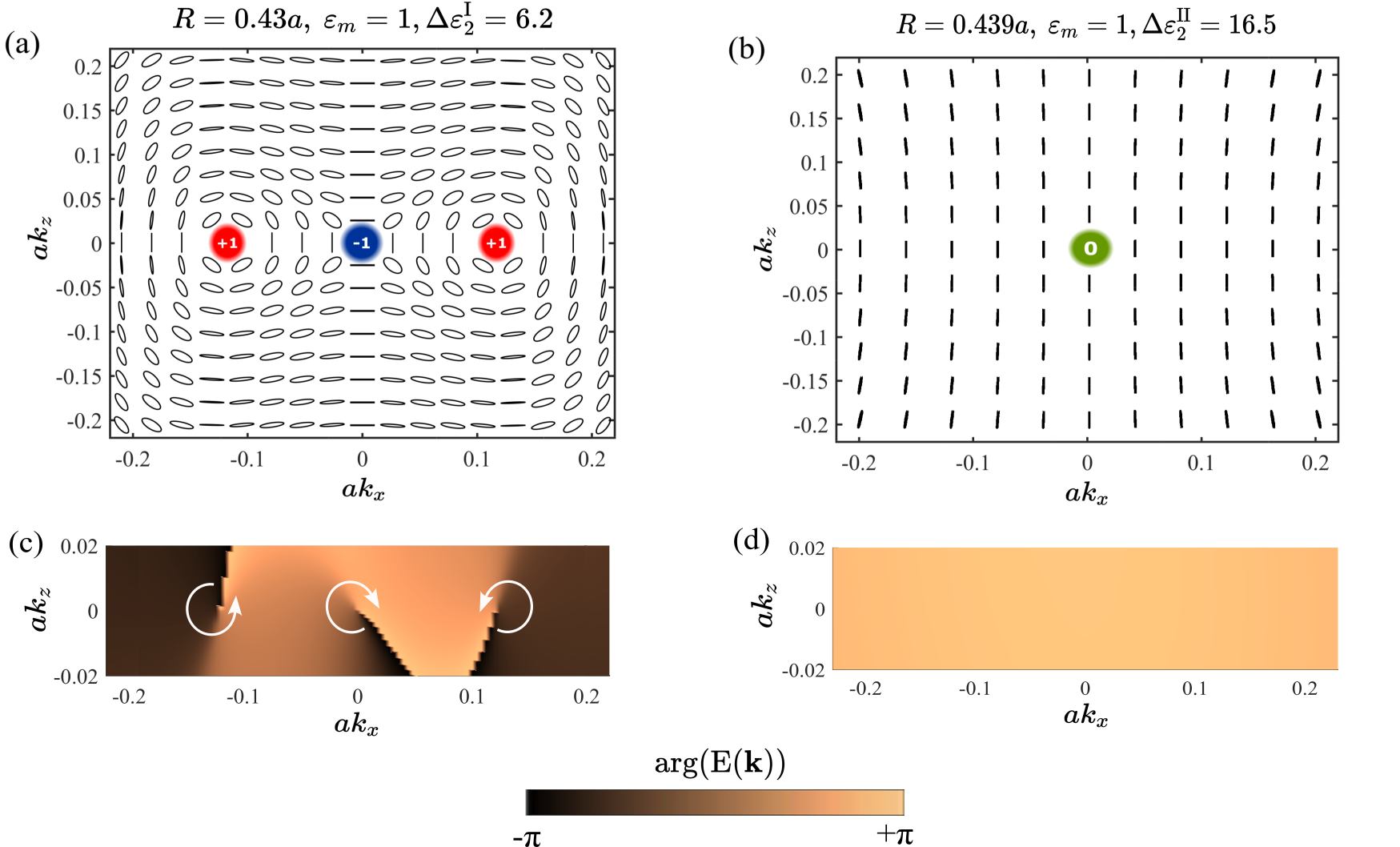}
          \caption{Topological charge value and far-field polarization ellipses corresponding to BICs' (a) transformation  and (b) annihilation processes. Panels (c-d) show the change of far-field polarization vector's phase $\text{arg}(\text{E}(\textbf{k}))$ in the parametric space. Circular arrows show the rotation direction of the polarization vector.}
          \label{fig:tg}
        \end{figure*}%
        
\begin{figure*}[htbp]
   \centering
    \includegraphics[width=0.6\textwidth]{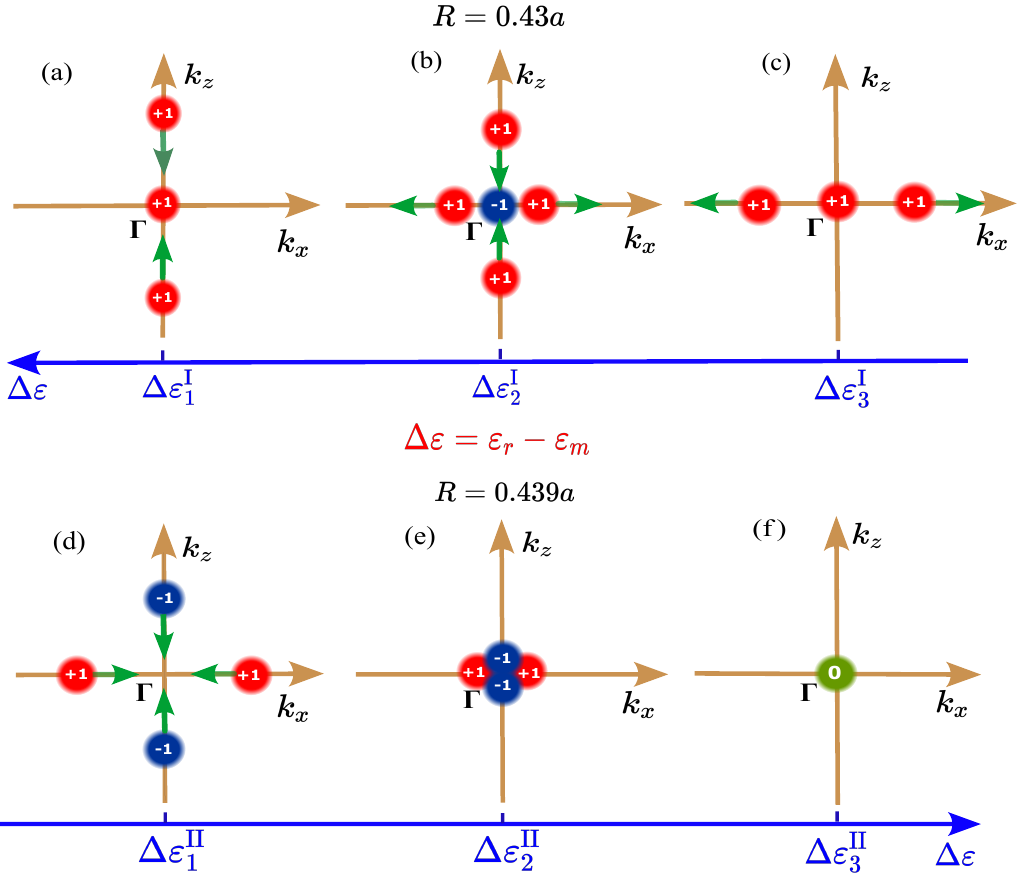}
          \caption{BIC migration and transformation 
in structures with (a-c) $R=0.43a$ and (d-f) $R=0.439a$. $\Delta\varepsilon=\varepsilon_r-\varepsilon_m$ is the optical contrast of the system, where $\varepsilon_r$ and $\varepsilon_m$ are the permittivity of the rods and surrounding medium,
          respectively.
          (a-c) For
          $\varepsilon_r=15$, $\varepsilon_m$ changes, and the critical values are$:$ $\Delta\varepsilon^{\textbf{I}}_1=14$, $\Delta\varepsilon^{\textbf{I}}_2=12.92$, and $\Delta\varepsilon^{\textbf{I}}_3=11.9$. 
          (d-f) For $\varepsilon_r=17.5$, $\varepsilon_m$ changes $:$ $\Delta\varepsilon^{\textbf{II}}_1=16.27$, $\Delta\varepsilon^{\textbf{II}}_2=16.47$, and $\Delta\varepsilon^{\textbf{II}}_3=16.5$. Upper index of $\Delta\varepsilon$ corresponds to the mode labels from Fig.~\ref{fig:structure} (b,c).
          Green arrows indicate the direction of BICs' migration.}
          \label{fig:pross}
        \end{figure*}
        
\begin{figure*}[htbp]
\centering
\includegraphics[width=0.9\linewidth]{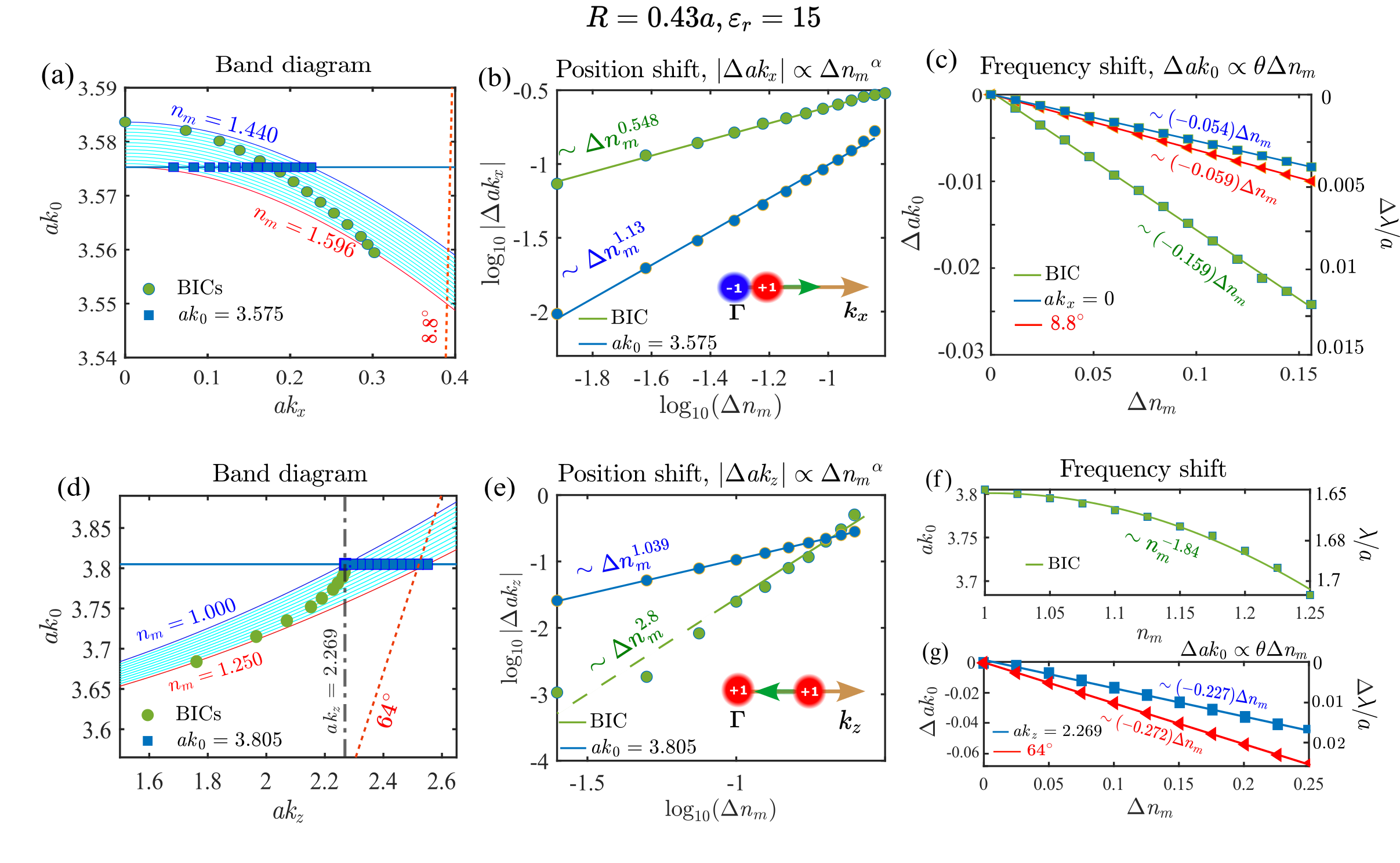}
          \caption{Frequency shift and BIC migration along (a-c) $k_x$ and (d-f) $k_z$ for $R = 0.43a$. $\Delta a k_0 = ak - ak_0$ and $\Delta n_m = n_m - n_1$ 
are the normalized frequency and refractive index changes, respectively. $n_m$ is within the range of 1.440–1.596 (step 0.012) for $k_x$ and 1.000–1.250 (step 0.025) for $k_z$. (b,e): Band shifts at fixed $ak_0$; (c,g): Frequency shifts at fixed $k_x$, $k_z$, and incidence angle. (b): $\alpha$[BIC(${k_x}$)]=$0.548\pm0.007$, $\alpha$[Band(${k_x}$)]=$1.31\pm0.02$; (e): $\alpha$[BIC(${k_z}$)]=$2.8\pm0.2$, $\alpha$[Band(${k_z}$)]=$1.039\pm0.005$. (c): $\theta$[BIC(${k_x}$)]=$-0.159\pm0.001$, $\theta$[Band(${k_x}$)]=$-0.054\pm0.001$, $\theta(8.8^\circ)$=$-0.059\pm0.001$; (f): $\theta$[BIC(${k_z}$)]=$-1.84\pm0.22$; (g): $\theta$[Band(${k_z}$)]=$-0.227\pm0.001$, $\theta(64^\circ)$=$-0.272\pm0.001$. (c,g) 
include $\Delta\lambda/a$ on the right $y$-axis. Insets in (b,e) show the BIC migration in the $k$-space.
          }         
          \label{fig:r0.43}
        \end{figure*}

We should note that the optical contrast can also be controlled by varying the permittivity of the rods. However, we found that resonance positions are more sensitive to changes in the medium's refractive index than to changes in the rod index.
The reason is, an increase in the rod permittivity reduces the fraction of field energy in the analyte, which negatively affects the angular sensitivity of the structure (see Section III of  Supplement 1.)

Figure~\ref{fig:tg} shows vortexes in the polarization directions of far-field radiation, while Fig.~\ref{fig:pross} presents the transformation of BICs with varying optical contrast. The vortexes carry conserved and quantized topological charges, defined by the winding number of the polarization vectors, which ensure their robustness and govern their generation, evolution, and annihilation~\cite{zhen2014topological}. The one-dimensional periodicity of the considered photonic structure restricts the absolute value of the topological charge to 1 (see Supplementary information at Ref.~\cite{zhen2014topological}).  
Both symmetry-protected and tunable off-$\Gamma$ BICs have nontrivial topological charges equal to $\pm 1$. In contrast, at-$\Gamma$ BIC shown in Fig.~\ref{fig:tg}(b) possesses a trivial topological charge. Consequently, it is not topologically protected and emerges accidentally due to the optimization of the structure's parameters and its optical contrast (see Section I of the  Supplement 1). In contrast to topologically protected BICs, if the parameters are detuned, at-$\Gamma$ BIC immediately transforms into a leaky state.

Note that, as shown in Fig.~\ref{fig:tg}(a), near the BICs, the polarization deviates from linear one in the directions where $ak_x \neq 0$ and $ak_z \neq 0$. In these cases, the topological charge definition given by Eq.~\eqref{Tcharge} becomes inconsistent~\cite{bulgakov2017bound}. In contrast, for lines with 
 either $ ak_x = 0 $ or $ ak_z = 0 $, the polarization remains linear~\cite{bulgakov2017bound}---a behavior that aligns with the observations in Fig.~\ref{fig:tg}(a,b) and results presented in Fig.~S2 of the  Supplement 1. Therefore, to accurately define the topological charge, we trace the phase evolution of the polarization vector, as illustrated in Figs.~\ref{fig:tg}(c,d) and the corresponding panels in Fig.~S2. Additionally, we found that the ellipticity of the polarization is more pronounced in the high-contrast structure (see Fig.~\ref{fig:tg}(b) and Fig.~S2 in the  Supplement 1), as polarization is sensitive to variations in the optical parameters.

Undoubtedly, the ability to tune BIC evolution through varying optical contrast opens up new possibilities for refractive index sensing, for example, of chemical and biological liquids~\cite{maksimov2022enhanced,jing2023high, hou2025high}. To estimate the sensitivity, we further studied both spectral $\Delta ak_0$ and angular $\Delta ak_{(x,z)}$ shifts of BICs for the 
the surrounding medium index within the range $1\leq n_m< 2$, which covers the available materials such as silicon oxide~\cite{kischkat2012mid}, water~\cite{kedenburg2012linear}, polymers and dyes~\cite{bohm2004tuning}, and biochemical liquids~\cite{peng2023optical}. The migration process is schematically shown in Fig.~\ref{fig:pross}. Namely, for $R=0.43a$, at-$\Gamma $ BIC decomposes into three BICs at $n_m=1.442$, i.e., $\Delta\varepsilon^{{\textbf{I}}}_2$ = 12.92. In a similar manner, $ak_{z}$-BICs annihilate at $\Gamma$ point when $n_m$ reaches 1.75,  see Fig.S2 in  Supplement 1. For the second structure with $R=0.439a$, a topologically trivial BIC emerges at the critical optical contrast $\Delta\varepsilon^{\textbf{II}}_3 = 16.5$, i.e.~for the rod and medium permittivities $\varepsilon_r = 17.5$  and $\varepsilon_m = 1$, respectively. Since this BIC is not topologically protected, it decomposes back into four BICs under 
any small perturbation of the refractive index of the medium. 
\begin{figure*}[htbp]
   \centering
 \includegraphics[width=0.9\linewidth] {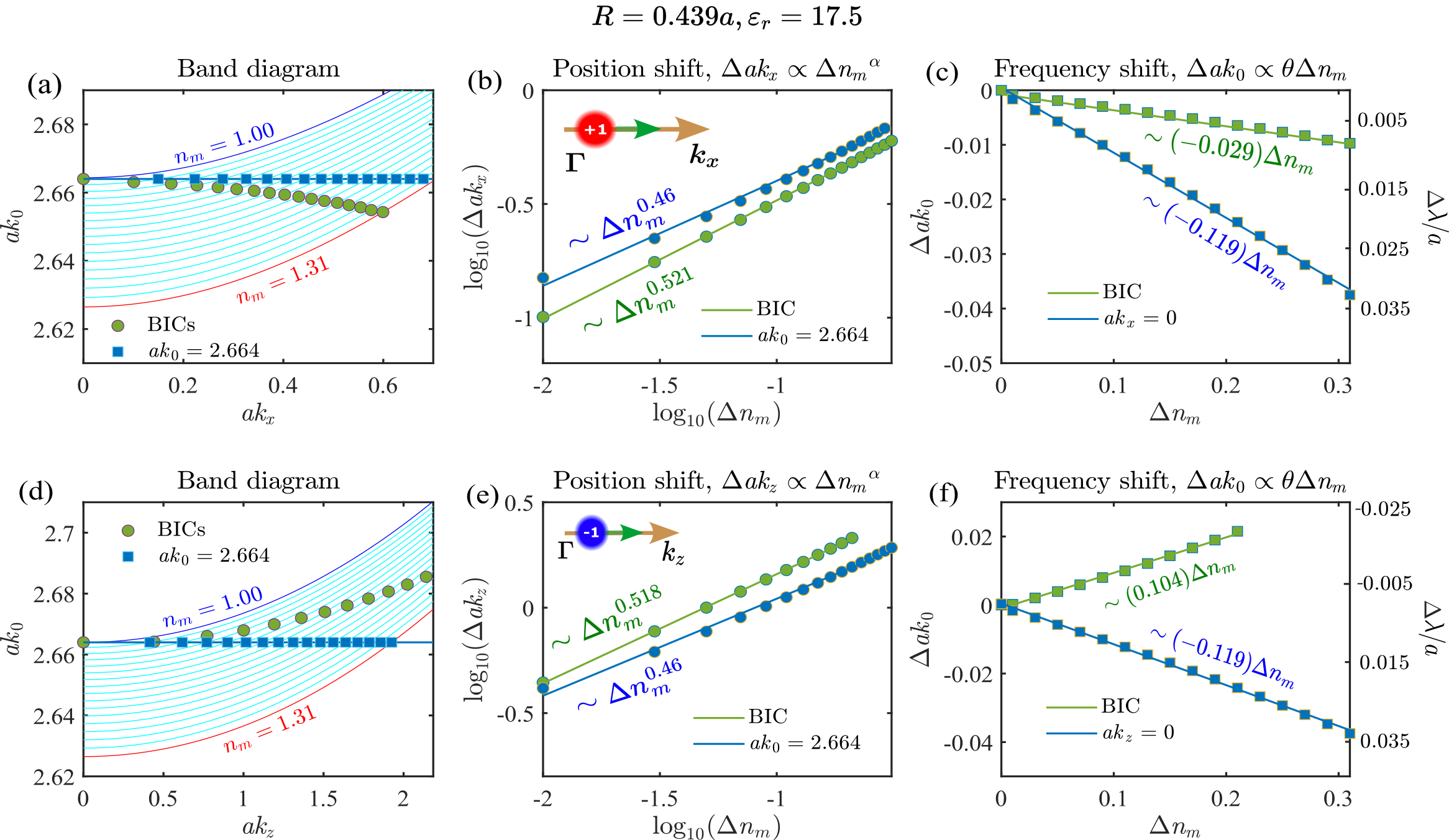}
\caption{Frequency shift and BIC migration along (a-c) $k_x$ and (d-f) $k_z$ for $R = 0.439a$, with refractive index $n_m$ varying from 1.00 to 1.31 in steps of 0.02. (b,e) Band shifts at fixed frequency; (c,f) frequency shifts at fixed $k_x$ and $k_z$. (b) $\alpha$[BIC(${k_x}$)] = $0.521 \pm 0.003$, $\alpha$[Band(${k_x}$)] = $0.46 \pm 0.01$; (e) $\alpha$[BIC(${k_z}$)] = $0.518 \pm 0.02$, $\alpha$[Band(${k_z}$)] = $0.46 \pm 0.01$. (c) $\theta$[BIC(${k_x}$)] = $(-2.94 \pm 0.04) \cdot 10^{-2}$, $\theta$[Band(${k_x}$)] = $-0.119 \pm 0.001$; (f) $\theta$[BIC(${k_z}$)] = $0.104 \pm 0.002$, $\theta$[Band(${k_z}$)] = $-0.119 \pm 0.001$. (c,f) include normalized wavelength change $\Delta\lambda/a$. Insets in (b,e) illustrate BIC migration direction in the $k$-space.}
          \label{fig:r_0439}
        \end{figure*}%

To predict BICs' displacement with respect to $\Delta n_m$, let us refer to perturbation theory. 
In the first order, the perturbed wavevector $k$ in the vicinity of the $\Gamma$ point is expressed as follows~\cite{yuan2017strong,maksimov2020refractive}:
\begin{equation}
        \label{Eq6}
        k=k_0(1+\eta\Delta\varepsilon_m-\kappa\Delta\phi^2+O(\Delta\varepsilon_m^2)-O(\Delta\phi_m^4)),
        \end{equation}
where $\varepsilon_m$ and $\phi=\theta-\theta_0$ are the permittivity of the medium and the change of the angle of incidence, respectively; the parameter $\kappa$ can be found from numerical or experimental data~\cite{maksimov2020refractive}, and the coefficient $\eta$ can be defined by overlap integrals~\cite{yuan2017strong}. 
One can prove from Eq.~\ref{Eq6} that the dependence of $\Delta\phi$ on $\Delta n_m$ is given by (see Section I of  Supplement 1 for details):

         \begin{equation}
        \label{Eq8}
        \Delta\phi=\sqrt{A+B\Delta n_m}.
        \end{equation}
here, $A$ is proportional to $\kappa$ associated with the band's curvature; $B$ is related to the coefficient $\eta$. Both $A$ and $B$ are normalized by the parameter $\kappa$. If we neglect the change of frequency, then we can conclude that in the vicinity of $\Gamma$ point, the BIC position $\Delta \phi$ is proportional to $\sqrt{\Delta n_m}$.

\textit{Change of BIC positions.}~Migration of BICs along the photonic band is shown in Figs.~\ref{fig:r0.43}(a) and~\ref{fig:r_0439}(a). In the case of $ R = 0.43a $, the BIC frequency decreases as
 $ n_m $ increases (see Fig.~\ref{fig:r0.43}).~The band diagram of mode I is parabolic, with a steep downward branch in the vicinity of the $\Gamma $ point, while BIC moves down along the curve with increasing $n_m $. Figure~\ref{fig:r0.43}(c) demonstrates that the frequency decreases linearly as $n_m$ increases. In contrast, Fig.~\ref{fig:r0.43}(d,f) indicates that the frequency shift of BIC along the $k_z$-axis ($ \text{BIC}(k_z) $) is nonlinear and increases for large values of $ n_m $. The angular shift of BIC along the $ k_x $ axis ($\text{BIC}(k_x)$) is proportional to $ \Delta k_x\simeq \Delta{n_m}^{0.5} $, while along the $ k_z $ axis, it scales as $ \Delta k_z\simeq \Delta{n_m}^3 $.

For $ R = 0.439a $, as $n_m $ increases, the dispersion curve shifts downward. However, the BICs moving in the $k_x$ and $k_z$ directions behave differently. Specifically, the frequency of BIC($k_x$) decreases, while the frequency of the BIC moving along the $k_z$ axis increases [Figs.~\ref{fig:r_0439}(a) and~\ref{fig:r_0439}(d)]. This behavior of the BIC($k_z$) contrasts with the conventional spectral shift, where increasing the refractive index typically reduces the
resonance frequency. Furthermore, in the vicinity of the $\Gamma$ point, these BICs show a relatively small spectral shift compared to the photonic band. The angular shift of BIC along both the $ k_x $ and $ k_z $ axes exhibits a dependence of $ \Delta k_x\simeq \Delta{n_m}^{0.5} $. Notably, $ \text{BIC}(k_z) $ is more sensitive to $ \Delta n_m $ compared to $ \text{BIC}(k_x)$.

The described features of BICs, indeed, are manifested in 
the dependence of BIC frequency shift on $ \Delta n_m $, shown in Figs.~\ref{fig:r0.43}(c,f) and ~\ref{fig:r_0439}(c, f). The slope angles of the fitted curves near the $ \Gamma $ point have small absolute values (see Section V of the  Supplement 1 for further details). The frequency shift is clearly associated with these slope angles. Therefore, as parameter $ A $ is also related to the frequency shift and relatively small compared to parameter $ B $ (see Sections I and VI of the  Supplement 1), we can conclude that in the vicinity of the $ \Gamma $ point, the BIC position in the $k$-space is proportional to $ \sqrt{\Delta n_m} $. In contrast, for the BIC migrating along $ k_z $ towards the $\Gamma$ point, the slope angle in Fig.~\ref{fig:r0.43}(f) is larger than for the BICs in the vicinity of the $ \Gamma $ point. 

\textit{Shift of photonic band.} Obviously, a change in $n_m$ also 
shifts the photonic band. To analyze the band shift for fixed values of the frequency and $k$-vector, we trace the dispersion curve of the modes for different values of the medium refractive index. For the 
mode I, we set $ak_0=3.575$ along $k_x$ and $ak_0=3.805$ along $k_z$ as initial values, see Fig.~\ref{fig:r0.43} (a,d). From Fig.~\ref{fig:r0.43}, we can conclude that the photonic band is less sensitive to changes in $n_m$ than BICs. For the mode II, as demonstrated in Fig.~\ref{fig:r_0439}(b,e), the angular shift of the photonic band is approximately proportional to the square root of $\Delta n_m$ irrespective of the direction (both for $k_x$ and $k_z$). To estimate the sensitivity of the mode II, we fix the values of $ak_0$ and $ak_x$  at $n_1=1$. Figure~\ref{fig:r_0439} shows that the band shift is slightly more sensitive to $\Delta n_m$ compared to
the BIC migration. It can be explained by the dispersion behavior 
in the vicinity of $\Gamma$ point: the dispersion curves of mode II in Figs.~\ref{fig:r_0439} (a,d) are less steep than those diagram 
of mode I [Fig.~\ref{fig:r0.43}(a)].  

Thus, in the rod array with  $R=0.43a$, BICs exhibit a higher angular and spectral sensitivity, i.e., a larger slope angle, compared to 
the photonic band [Fig.~\ref{fig:r0.43}].~Above, we considered fixed values of the wave vector projection with different values of the frequency, and vice versa. Here we set the angle of incidence $\phi$ to $8.8^\circ$ along the $k_x$-axis and $64^\circ$ along the $k_z$-axis. These values were chosen regarding the maximum incidence angle within the computed range of the band diagram shown in Fig.~\ref{fig:r0.43}. At these fixed $\phi$, we observe the frequency shift of the photonic band. The results in Fig.~\ref{fig:r0.43}(c,g) reveal that increasing $\phi$ enhances the spectral shift of the band. Although varying the angle of incidence allows  scaling the photonic band shift, its spectral sensitivity remains lower than that of the BICs. Next, we will discuss the exact values of angular and spectral sensitivities in more detail.

\section{Sensitivity}
\begin{figure}[htbp]
          \centering
\includegraphics[width=0.9\linewidth]{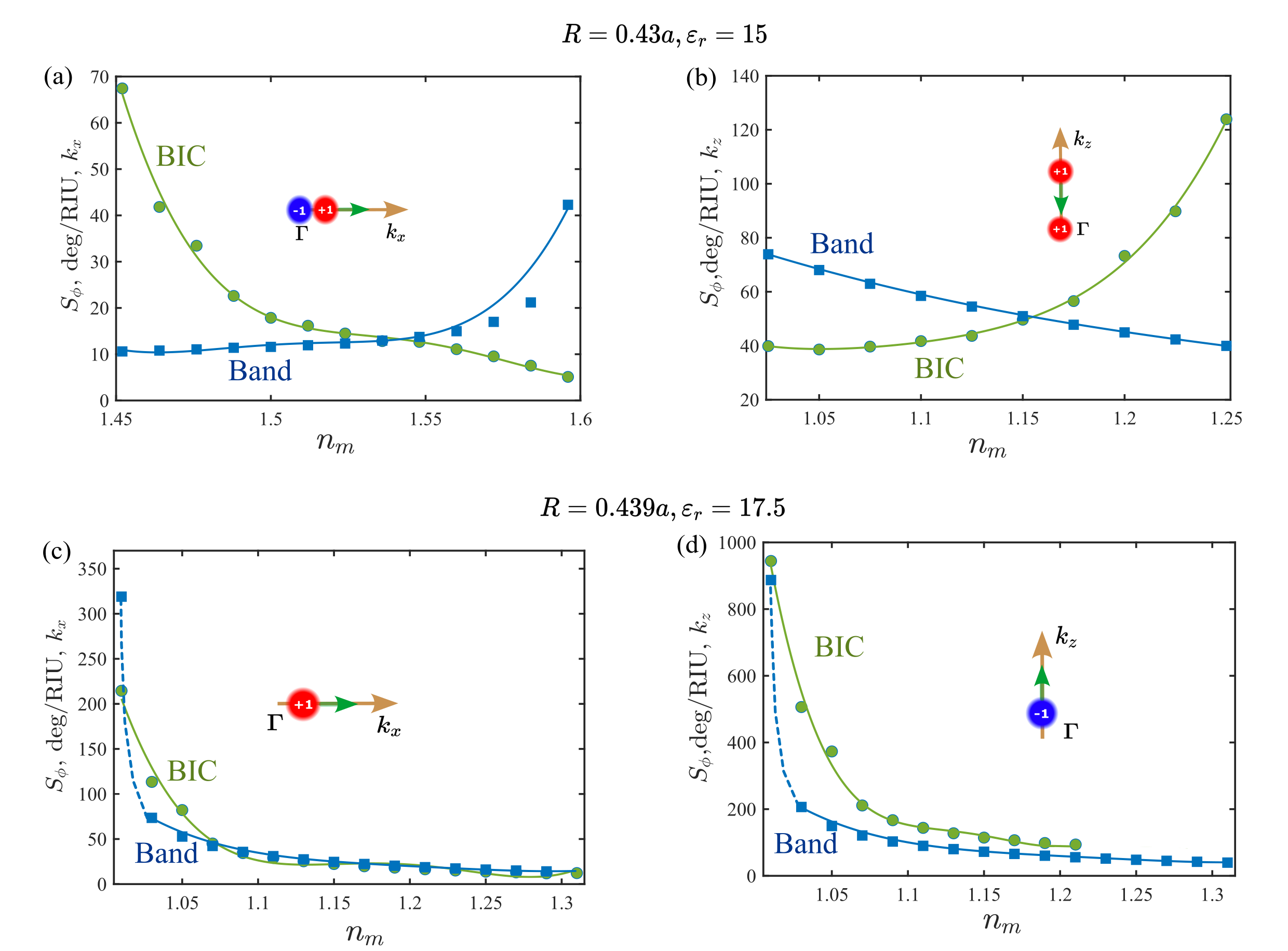}
          \caption{Angular sensitivity of the BIC and band for mode I (panels a, b) and mode II (panels c, d). 
          In panels (c, d), the first data point of the band exhibits high sensitivity and was excluded from the fitting process.} 
          \label{fig:Sens_043}
        \end{figure}

\begin{figure}[htbp]
          \centering
\includegraphics[width=0.6\linewidth]{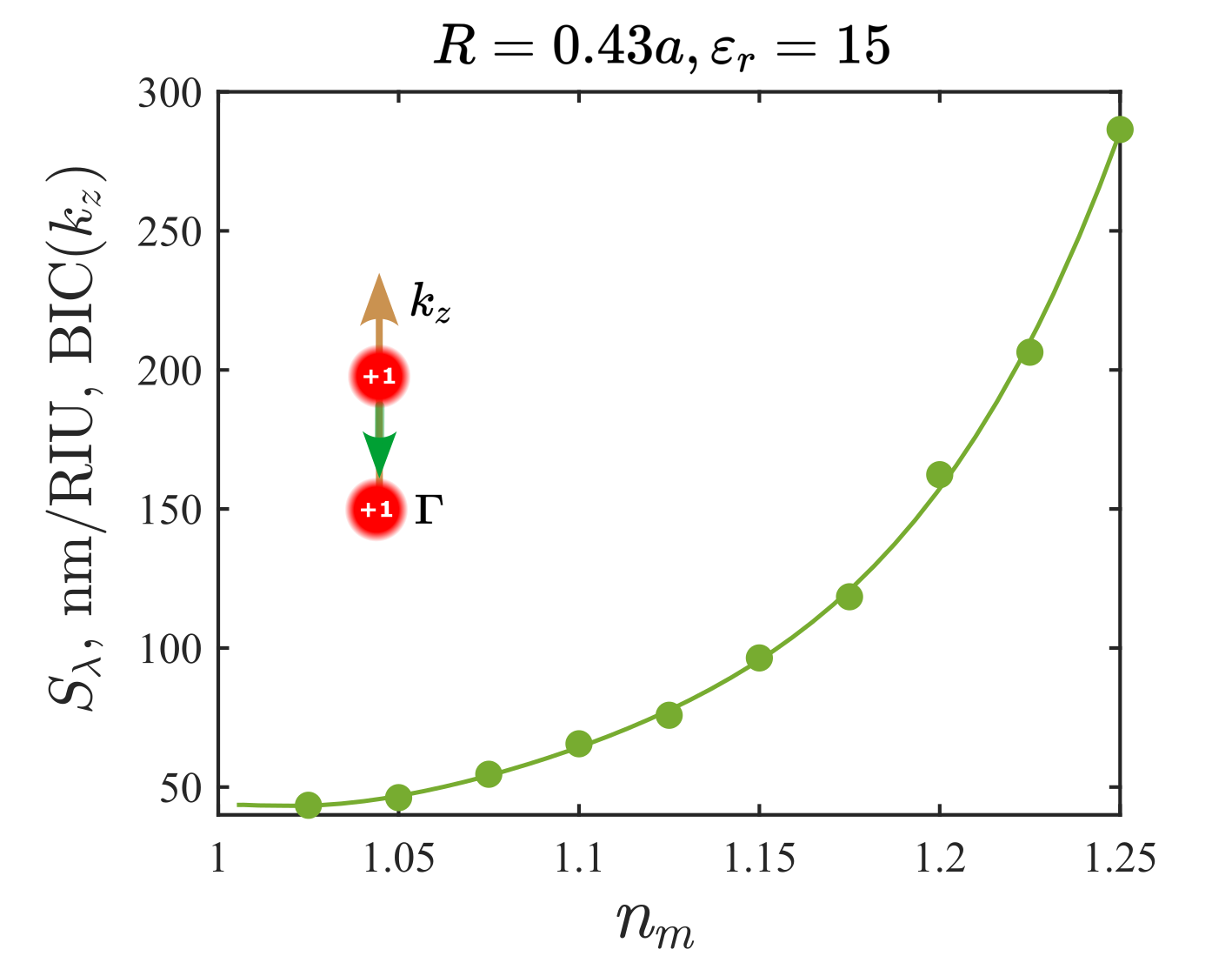}
          \caption{Spectral sensitivity of BIC that migrate to $\Gamma$ point. This migration is shown schematically in the inset.} 
          \label{fig:Sens_spec}
        \end{figure}

As previously discussed, the migration of BICs away from the $\Gamma$ point demonstrates low spectral sensitivity. However, their angular shift is proportional to the square root of the variation in the medium’s refractive index. Here we calculate the values of the angular and spectral sensitivities for given values of $n_m$. Results show that the spectral sensitivity of BICs near the $\Gamma$ point reaches $\sim{50}$ nm/RIU. On the other hand, as mentioned in the previous section, BICs exhibit a significant angular shift. Specifically, for a structure with $R=0.439a$ and $n_m=1.1$, the change of the incident angle achieves $7$ degree 
along the $k_x$ axis and $31$ degree
 along the $k_z$ axis. For the array with $R=0.43a$ at given $n_m$, the maximum change of the angle of incidence is $4$ degree. 
The obtained values of the angular sensitivities are shown in Fig.~\ref{fig:Sens_043}. For mode II, the angular sensitivity of the BIC($k_z$) for $n_m\sim1.01$ tends to 1000 deg/RIU, while with an increase of $n_m$, it dramatically drops. One of the reasons for such behavior is, at $n_m\lesssim 1.01$, BICs transform from topologically non-protected into topologically protected ones. Moreover, as we discussed in the previous section, the angular shift of BICs near the $\Gamma$ point also depends on the shape of the photonic band. Interestingly, the mode I BIC($k_z$) has an unpredictable sensitivity for $n_m\gtrsim1.15$. Indeed, as Fig.~\ref{fig:r0.43}(d) shows, the values of sensitivities at the last two points, i.e., $\Delta n_m=1.250-1.225=0.025$ are equal to $S_\phi=124$ deg/RIU, $S_\lambda=285.5$ nm/RIU. Furthermore, the spectral sensitivity of the mode I along $k_z$ is non-linear [Fig.~\ref{fig:Sens_spec}]. These unexpected values fall outside the scope of our theoretical predictions, as the migrations occur far from the $\Gamma$ point. Consequently, the theoretical explanation of the observed sensitivity behavior requires higher-order terms in the perturbation expansion of the $k$ vector.

Next, we compare the obtained magnitudes of the band and BIC sensitivities. In the case of  $R=0.439a$, the maximum angular sensitivity of the band reaches $10^{3}$ deg/RIU, while its maximum spectral sensitivity is approximately equal to $50$ nm/RIU. Similar to
the BIC angular sensitivity for small values of $n_m$, the high sensitivity of the band can be understood in terms of the photonic band shape, as discussed in the previous section. In contrast, for the mode I, band's spectral and angular sensitivities are lower than those of BICs, and their maximum values reach 
$42$ deg/RIU and $13$ nm/RIU along the $k_x$ axis. The angular and spectral sensitivities of the band  along the $k_z$ axis are equal to $74$ deg/RIU and $57$ nm/RIU. More details about the BIC dynamics and band shift are available in  Section IV of the  Supplement 1. All the 
maximum sensitivities data are summarized in Table~\ref{new}.
Additionally,  we analyze the spectral sensitivity for a fixed angle of incidence $\phi$. For this, we consider the structure with $R=0.43a$. The angle of incidence was  $\phi=8.8$ degree along $k_x$ and $\phi=64$ degree along $k_z$. As shown in Table~\ref{new}, at $\phi=64$ degree along $k_z$, spectral sensitivity is $59$ nm/RIU, while for $\phi=8.8$ degree along $k_x$, it equals to $16$ nm/RIU. Indeed, as Fig.~\ref{fig:r0.43}(c,g) shows, the slope angle is slightly larger for BIC than for the band. The reason is, the
intersection line becomes more tilted for large values of $\phi$.  
        
        \begin{table*}
\caption{ Comparison of 
maximum and minimum spectral $S_{\lambda}^{\text{min}}$ (nm/RIU), $S_{\lambda}^{\text{max}}$ (nm/RIU) and angular  $S^{\text{min}}_{\phi}$ (deg/RIU) , $S^{\text{max}}_{\phi}$ (deg/RIU) sensitivities of BICs and bands.}
\label{new}
    \centering
    \begin{tabular}{c |c c |c c}
     
\hline
  Mode I, $R=0.43a$&\mbox{$S^{\text{min}}_{\lambda}$} &\mbox{$S^{\text{max}}_{\lambda}$ } 
 &\mbox{$S^{\text{min}}_{\phi}$} &\mbox{$S^{\text{max}}_{\phi}$ } \\
  \hline
  BIC($k_x$)& \multicolumn{2}{c|}{38} & 5.1&67.4\\
    Band($k_x$)
   & \multicolumn{2}{c|}{13}&10.6 &42\\
   $ \phi=8.8 (\deg)$
   & \multicolumn{2}{c|}{16} &-&-\\
 BIC($k_z$)
   & 43.2&285.5&40 &124\\
   Band($k_z$)
   & \multicolumn{2}{c|}{57}& 40&74\\
  $ \phi=64(\deg)$
   & \multicolumn{2}{c|}{59} &-&-\\

  \hline
    \hline
    Mode II, $R=0.439a$&\mbox{$S^{\text{min}}_{\lambda}$} &\mbox{$S^{\text{max}}_{\lambda}$ } 
 &\mbox{$S^{\text{min}}_{\phi}$} &\mbox{$S^{\text{max}}_{\phi}$ } \\
     \hline
    BIC($k_x$)& \multicolumn{2}{c|}{15}&13&215\\
    Band($k_x$)
   & \multicolumn{2}{c|}{50} &14.3&319\\
   
    BIC($k_z$)
  & \multicolumn{2}{c|}{42} &98&944\\
   Band($k_z$)
   & \multicolumn{2}{c|}{50} &40.33&887\\
\hline
    \end{tabular}
   \label{tab:my_label}
\end{table*}
  
To conclude, we found that the investigated structure is quite sensitive to 
changes in the angle of incidence of the beam. Moreover, for the array with $R=0.43a$, BIC($k_x$) is more sensitive to $\Delta n_m$ than the band. Thus, for the mode I, BIC migration is advantageous for detecting
the variation of the medium refractive index. 
In contrast, in the other structure with $R=0.439a$, the BICs and dispersion curve  demonstrate similar sensitivities. 
Therefore, both the band shift and BIC migration can be monitored 
to achieve a high sensitivity.
As Fig.~\ref{fig:Sens_043} shows, the angular sensitivity of BICs decreases as the refractive index of the medium increases. Consequently, there is a range of $n_m$ values for which BICs exhibit a high angular sensitivity. We have discussed one possible reason above. Interestingly, for the mode I, where the BIC begins to migrate from the edge of $k_z$ Brillouin zone, at certain values of $n_m$, sensitivity increases more significantly for the linear dependence than for the square-root one. A possible reason for this is the nature of the square root function. As Fig. S5(a) in Section VI of the  Supplement 1 shows, the square-root dependence is more effective for small changes in the medium's refractive index. However, as the medium's refractive index increases, the linear dependence becomes more dominant.~Furthermore, the spectral sensitivity in the vicinity of the $\Gamma$ point is comparable to those reported in Ref.~\cite{zhao2024multi}, while the maximum spectral sensitivity is four times higher. In addition, the obtained angular sensitivity is of the order of $~10^2$, which is comparable with the results for SPP sensors \cite{nisha2019sensitivity}. 
As a proof of concept, we examined a realistic periodic array of GaAs rods with a period of $a = 500$ nm, radius of $R = 0.439a$, and permittivity $\varepsilon_r = 12$. To analyze the angular shift in this array, we computed the reflectance maps for refractive indices $n_m = 1$, $n_m = 1.33$, and $n_m = 1.45$, demonstrating a significant angular sensitivity see section VI of  Supplement 1). 
We should note that, due to fundamental and practical limitations in real structures, these effects should be studied in more detail. Undoubtedly, the sensitivity can be improved by optimizing the structure, which is a task for further investigations.

\section{CONCLUSION}
In this work, we have qualitatively explained the migration of optical vortices associated with bound states in the continuum (BICs) due to changes in the system's optical contrast. Based on the well-known result of perturbation theory, we reveal the square-root dependence of BIC position in a rod array from changes in the refractive index of the surrounding medium. By comparing the sensitivities provided by BIC shifts and band shifts at a fixed frequency and angle of incidence, we can conclude that for the first mode of the array, the BIC enables higher sensitivities, and for the 
second mode, the BIC and dispersion curve have similar sensitivities to the variation of the medium refractive index. Until now, a square-root dependence was observed only for systems supporting exceptional points \cite{wiersig2020review}. Furthermore, we observed red and blue shifts in the spectral response due to an increase in the medium refractive index. This finding can be used for advancing dielectric detectors that utilize angular changes in optical vortices, as these shifts enable the detection of very small changes in the medium 
index. This opens up possibilities for new applications and research of vortex-driven
detection.

\section{Acknowledgements}
Ravshanjon N. thanks Sergei Gladyshev and Prof. 	Alexey Shcherbakov for insightful discussions. The authors appreciate Lydia Pogorelskaya's proofreading of the English manuscript.
\section{Funding}
This work was supported by the Ministry of Science and Higher Education of the Russian Federation (Project FSER-2025–0012). The theoretical studies were supported by the Russian Science Foundation (Project 23-72-10059) and Priority 2030 Federal Academic Leadership Program.
\nocite{*}

\bibliography{ARTICLE/refs}

\end{document}


\begin{center}
\Large\textbf{Supplemental Document for ``Polarization Vortex for Enhanced  Refractive Index Sensing''}
\end{center}
\author{} 

\section{Derivation of \texorpdfstring{$\Delta\phi$}{Delta Phi}} 
Here, we present the derivation of the change of incidence angle $\Delta\phi$ due to varying refractive index of the medium. 
Changes in the medium permittivity $\Delta\varepsilon_m$
can be written as:
\begin{equation}
    \Delta\varepsilon_m=n_m^2-n_1^2=(n_m-n_1)(n_m+n_1)\approx 2 \Delta n_m n_1
\end{equation}

where $n_1$ is the initial value of the medium refractive index, and $n_m$ is the refractive index of the added analyte. 
For our case,
${n_m}/{n_1}<2$,  and $\Delta n_m^2$  is relatively small compared to 
$\Delta n_m$. In Eq.~2 of the main text, $\Delta\phi$ is the 
change of the incident angle, for small values of which the asymptotic behavior vanishes. After substituting this change of permittivity, Eq.~2 of the main text can be rewritten as:
\begin{equation}
\label{e4}
    \frac{k}{k_0} \approx 1+2\eta\Delta n_m n_1-\kappa\Delta\phi^2.
    \end{equation}
  
From Eq.~\ref{e4}, one can find the equation for $\Delta\phi$, which has the following form:
\begin{equation}
\label{Eq12}
     \Delta\phi \approx \sqrt{\frac{k_0-k+2k_0\eta\Delta n_m n_1}{k_0\kappa}}.
    \end{equation}

One can simplify the Eq.~\ref{Eq12} as:
\begin{equation}
\label{Eq13}
\Delta\phi=\sqrt{A+B\Delta n_m}.
\end{equation}

\section{Merging and annihilation of BICs}
\begin{figure} [H]
    \centering
    \includegraphics[width=0.65\linewidth]{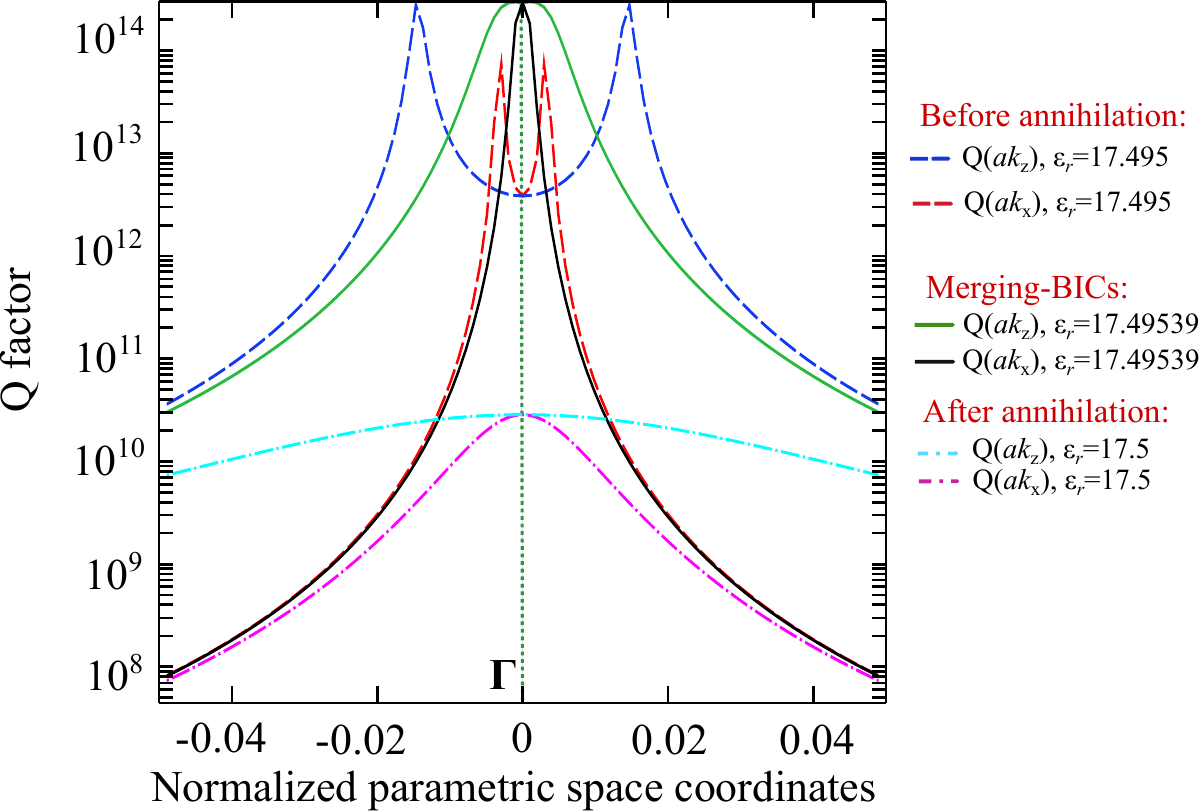}
    \caption{
    $Q$ factor before and after BIC annihilation in the vicinity of the $\Gamma$ point for the array with $R=0.439a$.  }
    \label{fig:sup_1}
\end{figure}

Figure~\ref{fig:sup_1} shows the $Q$ factor as a function of the 
rod permittivity 
for a structure with $R=0.439a$. 
An increase in $\varepsilon_{r}$ leads to the migration  off-$\Gamma$ BICs to the center of Brillouin zone.  When $\varepsilon_{r}=17.49539$, off-$\Gamma$ BICs are turned into the merging BICs~\cite{kang2022merging} with considerably enhanced the $Q$ factors. When $\varepsilon_{r}$ reaches 
critical value 17.5, the merging BICs annihilate. 
A further increase in the rod permittivity 
will reduce the $Q$ 
factor, 
indicating the decay of the topologically unprotected BIC. This phenomenon is associated with 
vanishing 
average value of the even BIC 
mode profile over the period~\cite{koshelev2021bound}. In contrast, the $Q$ factors of the off-$\Gamma$ and merging-BICs are infinitely high in the absence of radiative losses. 
However, in the simulation, their values are limited by numerical constraints.

\section{Polarization}
\begin{figure} [H]
    \centering
    \includegraphics[width=1\linewidth]{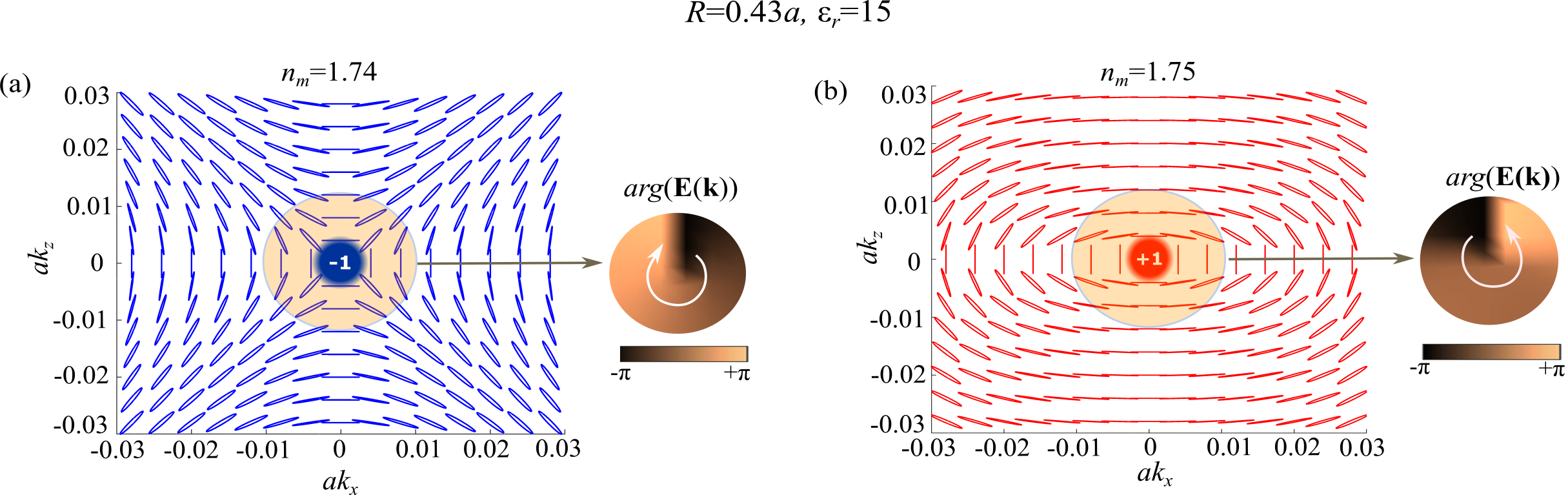}
    \caption{Far-field polarization ellipses in the vicinity of the $\Gamma$ point before (a) and after (b) the decomposition of BIC. Insets 
    show the phase  of the polarization vector $\text{E}(\textbf{k})$. Parameters of the structure are indicated in the titles of the 
    panels.}
    \label{fig:sup2}
\end{figure}
BICs also can be characterized in terms of topological charges. 
According to the article~\cite{zhen2014topological}, let us define  the topological charge by the expression:
\begin{equation}
\label{Tcharge}
    q=\frac{1}{2\pi}\oint_C \text{d} \textbf{k}\nabla \arg(\text{E}({\textbf{k}})),
\end{equation}
where the integral is taken along a closed path $C$ in the $k$-space that encircles the BIC in the counter-clockwise direction; $\text{E}({\textbf{k}})=c_{x}(\textbf{k}) + i c_{z}(\textbf{k})$, where $c_{x}(k)$ and $c_{z}(k)$ are 
the Cartesian components of the polarization vector. 

Figure~\ref{fig:sup2} shows the distributions of polarization ellipses in the far field in reciprocal space. The colored insets depict the phase profile of the polarization in the specified vicinity of the $\Gamma$ point. The white arrows indicate the direction of phase growth. The calculation results in Fig.~\ref{fig:sup2} indicate that for the structure with $R=0.43a$ and $\varepsilon_{r}=15$, the BICs at the $\Gamma$ point change their topological charge sign, when the refractive index of the medium reaches the critical value 1.75.

\section{Overlap integral}

 \begin{figure} [H]
    \centering
    \includegraphics[width=0.5\linewidth]{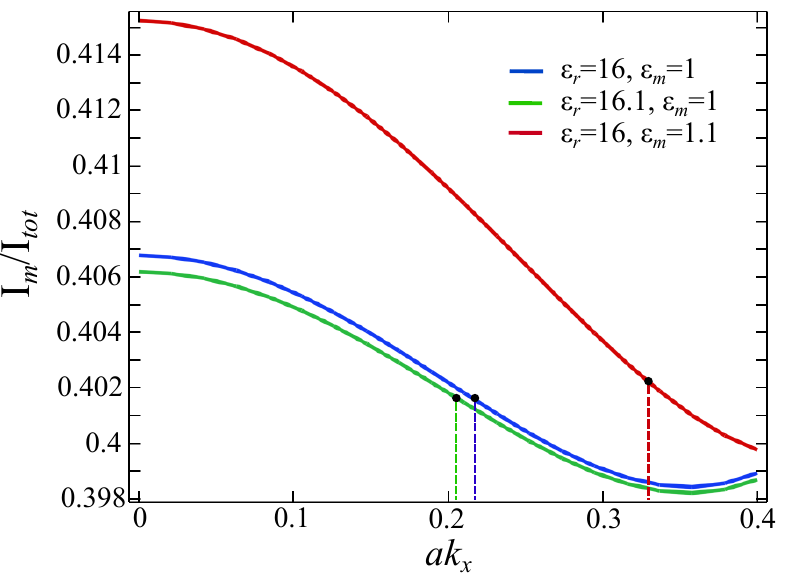}
    \caption{The overlap integral value normalized by the total field energy for the rod array 
    with $R=0.439a$. Dashed line indicates the BIC positions.}
    \label{fig:sup1}
\end{figure}
In the main text, we conclude that the structure is more sensitive to 
changes in the medium refractive index 
than the variation of rod permittivity.
An increase in the 
rod permittivity reduces 
the overlap of the field with the analyte, which makes the sensor less sensitive to the rod permittivity variation. To estimate this phenomenon, we calculate the values of the overlap integral (the amount of the field's energy in the analyte) for {an} array of rods with $R=0.439a$ and with initial 
permittivity $\varepsilon _{r}=16$.
When $\varepsilon_m$ increases to 1.1, the overlap integral grows by $1.34\%$, 
compare blue and red curves in Fig.~3. In contrast, the increase of rod permittivity by $\Delta\varepsilon_{r}=0.1$ leads to the decrease in the overlap integral by $0.1\%$. 
Undoubtedly, the value of overlap integral is important for the estimation of sensitivity of the system. Figure~\ref{fig:sup1} shows the 
the overlap integral calculated  
for different values of the optical parameters. As one can see, the 
overlap integral increases with increasing 
$\varepsilon_{m}$. From Fig.~\ref{fig:sup1}, where the dashed lines 
show the BIC positions, one can conclude that the system is more sensitive to the variation of the medium permittivity. 
 
\begin{table}[H]
\centering
        \caption{\label{tab:tables1}%
       Comparison of critical values of the system parameters for BIC annihilation and transformation}

        \begin{tabular}{llll}
      \hline\mbox{}&\mbox{$R/a$}&\mbox{$\varepsilon_r$}&\mbox{$n_m$}\\  
        \hline
    \mbox{\textcolor{blue}{Annihilation:}}&\mbox{}&\mbox{}&\mbox{}\\ 
 
        Article~\cite{bulgakov2017bound}  &0.44411&15&1\\
         This work &0.439&17.5&1\\
         This work &0.43&22.9&1\\
        \hline
         
           \mbox{\textcolor{blue}{Transformation:}}&\mbox{}&\mbox{}&\mbox{}\\ 
   Article~\cite{bulgakov2017bound}  &0.44&15&1\\
         This work &0.43&7.2&1\\
           This work &0.43&15&1.442\\
           \hline
       
                 \mbox{\textcolor{blue}{$k_z$-BICs annihilation:}}&\mbox{}&\mbox{}&\mbox{}\\ 
          
        Article~\cite{bulgakov2017bound}  &0.44&15&1\\
         This work &0.43&5.8&1\\
           This work &0.43&15&1.75\\
        
        \hline\end{tabular}
        \end{table}
  We should note that the values of critical optical contrast depend on 
   the parameters of the system. Table~\ref{tab:tables1} presents  the critical values, at which the annihilation and transformation processes are 
   observed. The differences between the obtained results and those of the article ~\cite{bulgakov2017bound} imply 
   that the critical parameters are related to 
   certain geometrical and optical parameters. Indeed, 
   changes of the structure parameters result in 
   complex behavior of BIC migration in the $k$ and parameter spaces~\cite{song2024evolution}. Thus, the values of critical parameters depend on 
   the structure characteristics and must be obtained in advance through numerical simulation. 
 \section{Band diagram}
\begin{figure} [ht!]
    \centering
    \includegraphics[width=0.55\linewidth]{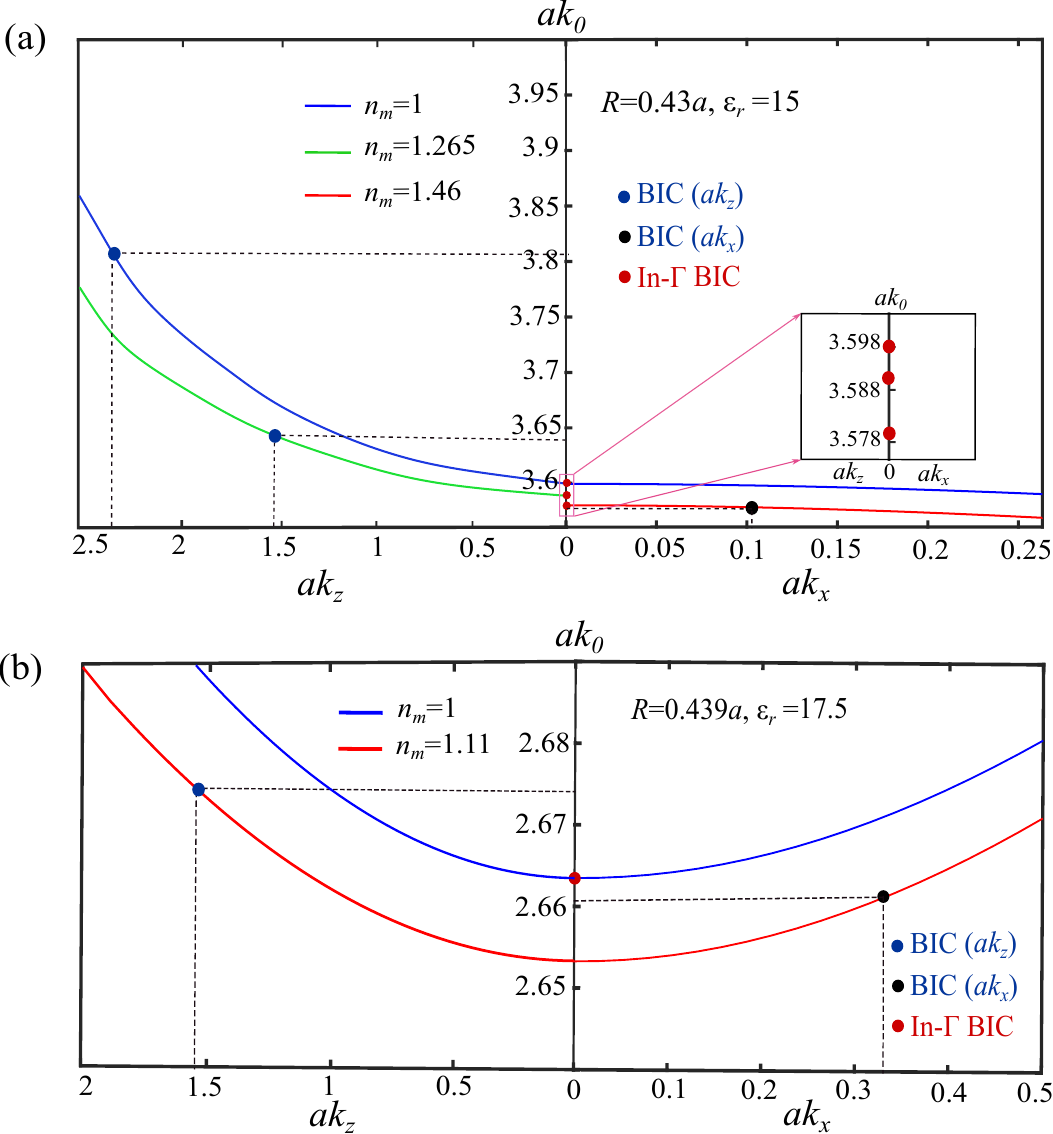}
    \caption{Dynamics of 
    BIC shifts along 
    the dispersion branch for different values of the refractive index of the medium. Dashed line indicates the BIC positions. 
    Parameters of the structures are shown in the insets.}
    \label{fig:sup3}
\end{figure}
Next, we compare the dynamics of BIC migration along 
the dispersion curve with increasing 
medium refractive index (Fig.~\ref{fig:sup3}).
 For the structure with $R=0.439a$, $\varepsilon_r=17.5$, and $\Delta n_m=0.11$, the absolute value of mode's normalized frequency shift is approximately equal to $0.01$.  For the structure with $R=0.43a$, $\varepsilon_r=15$, and $\Delta n_m=0.265$, this parameter takes value $0.01$ at the $\Gamma$ point and 
 $0.1$ in the vicinity of the Brillouin zone edge.  However, when the 
 refractive index of the medium changes to $0.46$, the spectral shift is equal to $0.02$. 
 For {a} 
structure with $R=0.439a$, $\varepsilon_r=17.5${,} and $\Delta n_m=0.11$, BICs 
change their coordinates to $0.33$ along $ak_x$ and to $1.51$ along $ak_z$ axis. {For} 
another structure with $\Delta n_m=0.265$, the value of $\Delta ak_z$ is about $0.75$. However, when $\Delta n_m$ increases to $0.46$, displacement along the $ak_{x}$ is approximately equal to $0.1$. These migrations are 
shown in Fig.~\ref{fig:sup3}.  

   
    \label{tab:my_label}

\section{Sensitivity}
\begin{figure} [ht!]
    \centering
\includegraphics[width=0.85\linewidth, angle=0]{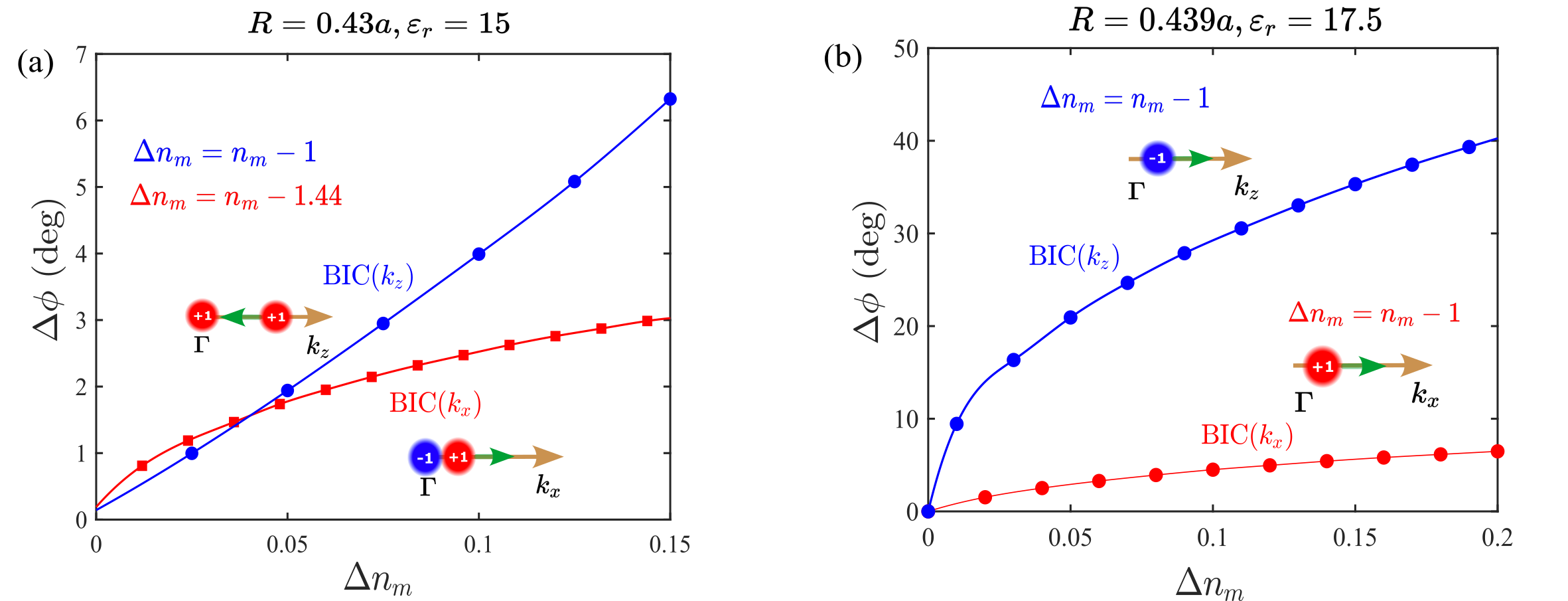}
    \caption{Change in the angle of incidence due to 
    altering 
    medium refractive index for the rod array 
    with $R=0.43a$ (mode I) and $R=0.439a$ (mode II). The arrows in the insets illustrate the migration processes. As the initial values of $\Delta n_m$ for the given scenarios are different, 
    the curve and the corresponding
    $n_m$ values are plotted in the same color for clarity.}
    \label{fig:sup3.4}
\end{figure}

      In the main text, we provide values for the maximum sensitivities of the BICs and the photonic band. 
      Here, we analyze the dependence 
      of the angular sensitivity on 
      the value of $\Delta n_m$. For the mode I, the optimal value of $n_m$ can be found in Fig.~\ref{fig:sup3.4} (a), where the nonlinear curve shows a greater change in $\phi$. It is worth noting that for the given scenarios, the initial values of $n_m$ are different and 
      obtained by the critical values of the optical contrast. To distinguish between different scenarios, 
      the curves and the initial values of $n_m$ are shown in the same color.
   

\begin{figure} [ht!]
    \centering
\includegraphics[width=0.85\linewidth, angle=0]{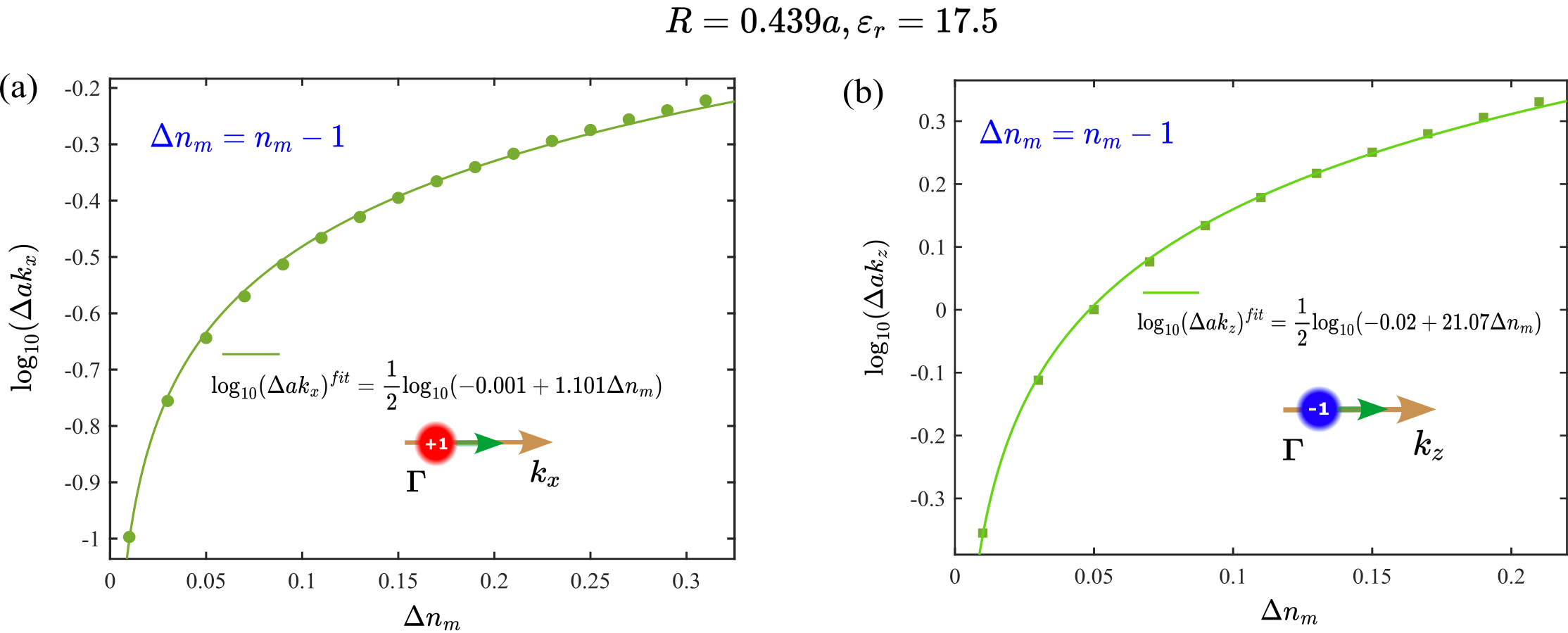}
    \caption{Approximation results of the equation $\Delta\phi=\sqrt{A+B\Delta n_m}$ for the mode II. Panel(a) 
    presents the results for $\Delta\phi\sim\Delta ak_x$, while the panel(b), 
    for $\Delta\phi\sim\Delta ak_z$. Insets show the direction of BIC migration.}
    \label{fig:sup3.42}
\end{figure}

   To analyze the sensitivity of BICs, we approximate the equation $\Delta\phi=\sqrt{A+B\Delta n_m}$ for the mode 
   II: in the case of BIC($k_x$), we replace $\phi$ with $ak_x$, while for BIC($k_z$), $\phi$, is represented as $ak_z$. The resulting expression is shown in Fig.~\ref{fig:sup3.42}. In both cases, the coefficient $A$ is relatively small, while $B$ is completely different. Therefore, $B$ affects the sensitivity of BICs more when $A$ tends to zero. More details 
   on these coefficients can be found in the main text.


\section{Reflectance}

\begin{figure*} [ht!]
    \centering
    \includegraphics[width=1\linewidth]{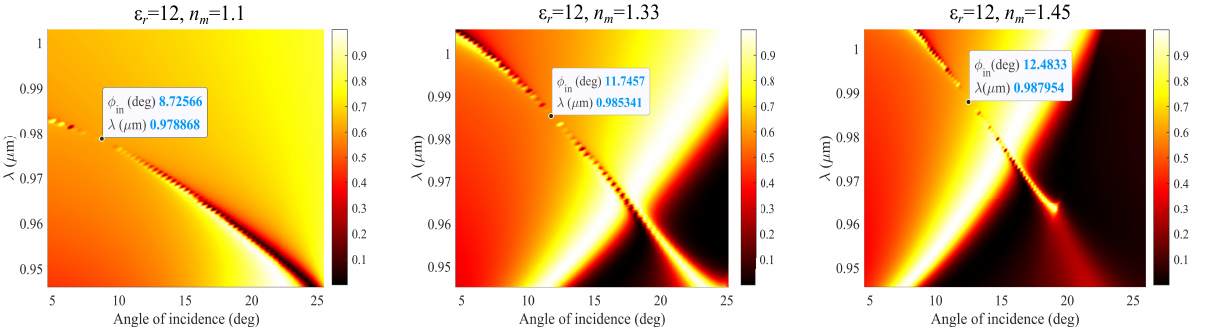}
    \caption{Reflectance spectra for the rod array 
    with $R=0.439a$. The indicated values are correspond to BICs. }
    \label{fig:sup4}
\end{figure*}

In addition to the investigated structures, we consider a periodic array with period $a=500$ nm, which consists of GaAs rods with radius $R=219.5$nm and 
permittivity $\varepsilon_{r}=12$. To analyze the sensitivity of the structure, we calculate the reflectance map for different values of the medium refractive index: $n_{m}=1$, $n_{m}=1.33$, and $n_{m}=1.45$. Reflectance maps were obtained by Fourier Modal Method (FMM)~\cite{spiridonov2023reformulated}. From Fig.~\ref{fig:sup4}, one can see that 
as $n_m$ increases from 1 to 1.45, 
the spectral shift is about 10~nm, while angle of incidence changes by 
3.76 degrees. 

